\documentstyle[12pt,a4]{article}

\def\by#1#2{{\displaystyle {#1}\over \displaystyle {#2}}}

\def\d{{\partial}}
\def\aleq{\raisebox{-0.4ex}{$\stackrel{<}{\sim}$}}
\def\ageq{\raisebox{-0.4ex}{$\stackrel{>}{\sim}$}}
\begin{document}

\begin{flushright}
IMSc/98/11/53 \\
13th Nov 1998 \\ [0.5cm]
\end{flushright}

\begin{center}
{\Large \bf Hints of higher twist effects in the slope of the proton
structure function } \\ [0.5cm]
{\it Rahul Basu\footnote{E-mail: rahul@imsc.ernet.in} and D.
Indumathi\/\footnote{E-mail: indu@imsc.ernet.in; On leave from Mehta
Research Institute, Allahabad 211 019}}, \\
The Institute of Mathematical Sciences, CIT Campus, Madras 600 113. \\
[1cm]
\end{center}

\noindent {\bf Abstract}: We critically analyse the data available on
the reduced cross-section in deeply inelastic $e\,p$ scattering from
the H1 collaboration at HERA. We use available data on the longitudinal
structure function to deduce the nature of $\d F_2/\d\ln Q^2$ at
different $Q^2$ for fixed values of $x$ near $x \sim 10^{-4}$. We
present the results in a manner which effectively isolates possible
higher twist effects in the structure function $F_2$.

\vspace{0.2cm}


\vspace{0.5cm}

\noindent Deeply inelastic scattering (DIS) at the  $e\,p$ collider at
HERA has provided precision data on the proton structure functions over
the last few years. Data on the slope of the structure function,
$F_2(x, Q^2)$, was presented for the first time last year
\cite{Zeusdf2,H1df2}. This showed a surprising dip in the slope, $\d
F_2/\d\ln Q^2$, below $x \sim 10^{-4}$ (see Fig.~1), although $F_2$
itself continued to show a rise towards smaller $x$ down to the
kinematical limit. Such a dip was not anticipated or predicted by
available parametrisations at that time \cite{Parm}. There have since
been intensive discussions on this effect using leading twist as well
as higher twist contributions within a perturbative framework.  In
fact, there now exist new parametrisations \cite{Grv98,Mrs98}, which
attempt to incorporate this effect, albeit in a leading twist
analysis.  The case for higher twist effects is not yet overwhelming
\cite{Ht} although they occur naturally in an operator product approach
to DIS \cite{Muta}. In this context, it is relevant to ask whether
there exist other data or methods which, when combined with existing
data on $\d F_2/\d\ln Q^2$, can clearly indicate the presence of higher
twist terms, which have long been poorly understood in DIS. This letter
addresses this issue. In particular, we show that the slope of the
reduced DIS cross-section, $\sigma_r$, defined in eq.~(1) below, is
very sensitive to the slope of $F_2$. Along with available data on the
longitudinal structure function, $F_L(x, Q^2)$, we show that the slope
of $\sigma_r$ yields information on the nature of the higher twist
content of $\d F_2/\d\ln Q^2$ and hence that of $F_2$.

Recently, the H1 collaboration \cite{H1sr} at HERA has measured the
reduced cross-section, $\sigma_r$,
$$
\sigma_r(x, Q^2) = \left [ F_2(x, Q^2) - \by{y^2}{Y_+} F_L(x, Q^2)
\right]~,
\eqno(1)
$$
which is a clear indicator of the size of the longitudinal structure
function, $F_L$, compared to that of $F_2$, in the measured deep
inelastic $e$\,$p$ cross-section. Here $y$ is determined by $Q^2 = s x
y$ with $Y_+ = 1 + (1-y)^2$, where $s$ is the square of the total cm
energy (which is constant at HERA) and the two kinematical variables,
$Q^2$ and $x$, are as usual the momentum transfer and Bjorken scaling
variable respectively.

The quantity $\sigma_r$ has been independently measured as a function
of both $x$ and $Q^2$.  Hence, the slope of $\sigma_r$ with respect to
either $x$ or $Q^2$, with the other variable kept fixed, can be
determined. Furthermore, since $y$ lies between 0 and 1, it is clear
that $F_L$ contributes significantly to $\sigma_r$ only at fairly large
values of $y$. For example, even when $y = 0.5$, the quantity
multiplying $F_L$ in $\sigma_r$ is only $f_2 \equiv y^2/Y_+ = 0.2$.
Finally, $F_L = 0$ exactly, at leading order, due to the Callan-Gross
relation and is non-zero only at next-to-leading order. Hence, $F_L$ is
suppressed (by at least one power of $\alpha_S$) compared to $F_2$. As
a consequence, at low $y$, $\sigma_r$ is essentially determined by
$F_2$; conversely, any study of $F_L$ must therefore be made in the
large $y$ region for maximum sensitivity.

The data on $\d F_2/\d\ln Q^2$ is averaged over $Q^2$, with the average
$\langle Q^2 \rangle$ increasing with increasing $x$ in such a way that
all the data corresponds essentially to low $y$ \aleq\ 0.3; hence, it
is relevant to ask whether the behaviour shown in Fig.~1  persists at
all $Q^2$. In short, what is the $Q^2$ dependence of $\d F_2/\d \ln Q^2$
at fixed $x$ values ?

We analyse the H1 $\sigma_r$ \cite{H1sr} data with a view to extracting
such a $Q^2$ dependence. In order to do this, we shall separately
analyse the small $y$ and large $y$\ $\sigma_r$ data. To begin with, we
recognise that the $Q^2$ dependences of $F_L$ and $\d F_2/\d\ln Q^2$
are related. This is because, from eq. (1), we have
$$
\begin{array}{rcl}
S_x \equiv \left. \by{\d\sigma_r}{\d\ln y}\right\vert_x & = & 
\by{\d F_2}{\d\ln Q^2} - f_1 F_L - f_2 \by{\d F_L}{\d \ln Q^2}~; \\
& & \\
S_q \equiv \left. \by{\d\sigma_r}{\d\ln y}\right\vert_{Q^2} & = & 
-\by{\d F_2} {\d\ln x} - f_1 F_L + f_2 \by{\d F_L}{\d \ln x}~. 
\end{array}
\eqno(2)
$$
Here the factors $f_2 = y^2/Y_+$ and $f_1 = 2 y^2 (2-y)/Y_+^2$\ are
significant only for large $y$. In particular, $f_1 \sim 1$ when $y =
0.7$; furthermore, $f_1 \sim 2 f_2$ over the entire $y$ range. It is
observed that $\sigma_r$ is a fairly linear function of $\ln y$ at all
$y$ as can be seen from Fig.~2, where $\sigma_r$ has been plotted as a
function of $y$ for some selected $x$ values ranging from $x = 10^{-4}$
to $10^{-2}$. Hence the slope $S_x$ can be obtained (at these different
$x$ values) from straight line fits to the $\sigma_r$ data.

\paragraph{The small $y$ data}: The $y$--derivative of $\sigma_r$
(whether at constant $x$ or at constant $Q^2$) is insensitive to $F_L$
or its slope when $y$ is small.  In other words, the behaviour of $S_x$
at low $y$ values (and hence low $Q^2$ for a given $x$ value) directly
constrains the slope of $F_2$. The resulting slopes, $\d F_2 /\d\ln
Q^2$, are shown in comparison with those extracted differently by ZEUS
\cite{Zeusdf2} in Fig.~3. Note that in our calculation we have obtained
the slope at fixed $Q^2$, equal to the average $Q^2$ of the
corresponding ZEUS data for different values of $x$. All the points
have $y$ \aleq\ 0.3; in fact the large $x$ ($x > 10^{-3}$) data have
$y$ \aleq\ 0.15. The error bars in our extraction of the slope are due
only to the errors arising from the straight--line fit to the
$\sigma_r$ data while the ZEUS data include both statistical and
systematic errors. We see that there is good agreement between the two
data sets, leading us to conclude that $f_1 F_L \ll \d F_2/\d \ln Q^2$,
so that $S_x$ in eq.~(2) in indeed saturated by $\d F_2/\d\ln Q^2$ at
small $y$. The specific values of the slope, $\d F_2/\d\ln Q^2$ for
certain $x$ and $Q^2$ values have been shown in Table 1 labelled as
``small $y$'' data; we shall need this later on in our analysis.

\begin{table}
\centering
\begin{tabular}{|c|c|c|c|} \hline
\multicolumn{4}{|c|}{Small $y$ data} \\
$x$ & $Q^2$ & $y$ & $\d F_2/\d\ln Q^2$ ($= \d \sigma_r /\d\ln y\vert_x$)
\\ \hline
1.4 $10^{-4}$ & 3.85 & 0.305 & $0.371 \pm 0.017$  \\
2.4 $10^{-4}$ & 5.24 & 0.242 & $0.376 \pm 0.010$  \\
4.0 $10^{-4}$ & 8.76 & 0.243 & $0.345 \pm 0.005$  \\
6.2 $10^{-4}$ & 10.63 & 0.190 & $0.338 \pm 0.010$ \\ \hline
\multicolumn{4}{|c|}{Large $y$ data} \\
$x$ & $Q^2$ & $F_L$ & $\d F_2/\d\ln Q^2$ ($= \d \sigma_r
/\d\ln y\vert_x + f_1 F_L$) \\ \hline
1.4 $10^{-4}$ & 8.84 & $0.51 \pm 0.3$ & $0.921 \pm 0.32$ \\
2.4 $10^{-4}$ & 15.2 & $0.35 \pm 0.3$ & $0.756 \pm 0.32$ \\
4.0 $10^{-4}$ & 25.3 & $0.33 \pm 0.27$ & $0.675 \pm 0.29$ \\
6.2 $10^{-4}$ & 34.7 & $0.39 \pm 0.28$ & $0.758 \pm 0.30$ \\ \hline
\end{tabular}
\caption{The slope $\d F_2/\d\ln Q^2$ evaluated using eq.~2 for
different
(small) $x$ values for the small and large $y$ $(= 0.7)$ data. Note
that $\d \sigma_r /\d\ln y\vert_x$ is constant for all $y$ (see Fig.~2)
and directly equals $\d F_2/\d\ln Q^2$ at small $y$. The data on $F_L$
are taken from the H1 Collab \protect{\cite{H1fl}}. }
\end{table}

\paragraph{The large $y$ data}:  In the large $y$ region, the $F_L$
contribution can no longer be neglected. We therefore use the value of
$F_L$ determined by the H1 collaboration \cite{H1fl} at various $x$
values, for fixed $Q^2$ (corresponding to large $y = 0.7$) to analyse
the large $y$ $S_x$ data.  Hence the large $y$ analysis will be
restricted to $y = 0.7$.  Since $F_L$ has contributions only at NLO,
its slope ($\d F_L/\d\ln Q^2$) is rather small; this term is further
suppressed by $f_2$ ($\sim 0.45$ at $y = 0.7$). Hence, we neglect the
contribution of the slope of $F_L$ in what follows. We shall comment on
the validity of this approximation later on. Then $S_x$ in eq.~(2),
along with the data on $F_L$, yields a value for $\d F_2/\d \ln
Q^2\vert_x$ at large $y = 0.7$ within this approximation. Sparse data
is available for $F_L$ in the region $1\times 10^{-4}$
\aleq\ $x$\ \aleq\ $6\times 10^{-4}$; hence we can extract $\d
F_2/\d\ln Q^2$ only at these $x$ values. These values (which we refer
to as ``large $y$'' data) at different $Q^2$, but {\it at the same $x$
values as the ``small $y$'' data}, are shown in Table 1.  The ``large
$y$'' sample obviously corresponds to a larger $Q^2$ than the ``small
$y$'' data at a given $x$; however, note that the average $Q^2$ in the
sample we have analysed increases with $x$. The large error bars (much
larger than that of the small $y$ data) are essentially due to large
errors in the $F_L$ data.

We have therefore extracted $\d F_2/\d\ln Q^2$ as a
function\footnote{The
number of $x$ values is restricted by $F_L$ data. There exists more data
on $F_L$ from H1 \cite{H1flnew} that has been extracted using data on
$S_q$.  This data is consistent with the existing ones in the region of
overlap, but we shall not use them here since the data are at slightly
different $y$ values.} of $x$ in the range $1$--$6 \times 10^{-4}$\ .
These values are listed in Table 1. At each $x$ value,
we have obtained $\d F_2/\d\ln Q^2$ at two different $Q^2$ values
corresponding to small and large $y$ data as discussed above. Note that
in all cases $Q^2$ \ageq\ $4$ GeV${}^2$, which corresponds to a fairly
stable perturbative regime.

We use the NLO GRV (1994) \cite{Grv94}, GRV (1998) \cite{Grv98} and MRS
(1998) \cite{Mrs98} parametrisations as typical indicators of the
theoretical expectation based on purely twist--two perturbative DGLAP
\cite{Dglap} evolution equations.  These predict a primarily
logarithmic dependence of $F_2$ on $Q^2$ (along with a small $1/Q^2$
piece from the heavy quark contributions). This implies that the slope,
$\d F_2/\d\ln Q^2$, at fixed $x$ is essentially flat, with a small
(positive) slope due to the charm quark contribution.

In Fig.~4 we show the extracted large and small $y$ data samples as a
function of $x$ along with the NLO fits from the different
parametrisation sets at the same $(x, Q^2)$ values as the data.  Since
$\d F_2/\d\ln Q^2$ increases with $Q^2$, the upper points correspond to
the large $y$ (and hence larger $Q^2$) sample (See Table 1). The bigger
error bars on these points are due to the larger uncertainties in
$F_L$.  The errors on the small $y$ sample arise from the errors on our
fits to the slope of $\sigma_r$ and are much smaller. (The error bars
on the $\sigma_r$ data are very small except at the edges of the
kinematically accessible regions; we have included them in the fits to
the slopes but not in the error estimates of the slope). We have also
plotted the values for $\d F_2/\d\ln Q^2$ (averaged over $Q^2$)
obtained by ZEUS \cite{Zeusdf2} at the same $x$ values, at the same
average $Q^2$ as the small $y$ data, for comparison.

We see that all parametrisations are consistent (to within 1$\sigma$)
with the large $y$ data.  In the case of the small $y$ data, which also
corresponds to the smaller $Q^2$ sample, the requirements are more
stringent due to the smaller error bars. The 1998 fits are in general a
better fit than the 1994 one. However, it is clear that if the error
bars decrease due to more and better data being made available, while
the values remain near the current central ones, a pure change in
normalisation will not suffice to fit both the data sets.  This is
because the central values of the two data sets differ by more than
100\%, while the parametrisations differ by less than 50\%. It
therefore appears that the small $y$ (or equivalently, small $Q^2$)
data are suppressed relatively more than the large $y$ data. In
particular, the separation between the data at the smallest $x$ value
($= 1.4\times 10^{-4}$) clearly indicates a substantial evolution of
$\d F_2/\d\ln Q^2$ from $Q^2 \sim 4$ to $9$ GeV${}^2$ (See Table 1).
Such a behaviour can be attributed to higher twist effects. To
elaborate further, $F_2$ can be expressed in terms of a leading twist
and higher twist part, as $$ F_2 = C_{LT} \, \ln Q^2 +
\by{C_{HT}}{Q^2}~, $$ where the coefficients are in general functions
of $x$ and we have included only an additional twist--4 piece. This
results in a $Q^2$ dependence of $\d F_2/\d\ln Q^2$ of the form $$
\by{\d F_2}{\d\ln Q^2} = C_{LT} - \by{C_{HT}}{Q^2}~.  \eqno(3) $$ That
$\d F_2/\d\ln Q^2$ evaluated from the parametrisations at a given $x$
decreases with $Q^2$ is due to threshold effects from terms like
$m_c^2/Q^2$ occurring in the charm contribution. Any further
suppression of $\d F_2/\d\ln Q^2$ with decreasing $Q^2$ then arises due
to higher twist effects coming from the light quark sector and can
explain the trend of the data shown in Fig.~4. Such an effect will be
particularly visible for smaller $Q^2$ data, such as that corresponding
to the smallest $x$ value in Fig.~4: this results in the small $y$ data
being suppressed more than the large $y$ data. This explains the
observed agreement of the large $y$ data with existing twist--2
parametrisations while the small $y$ data at the same $x$ value
disagree by as much as $2\sigma$. This disagreement disappears with
increasing $x$ for both large $y$ and small $y$ data since $Q^2$ also
increases with $x$. It is rather surprising that higher twist effects
are visible at such seemingly large $Q^2$. However, it must be
remembered that this is the first time that the slope of $F_2$ has been
so precisely measured.  It thus appears that the $Q^2$ dependence of
the slope, $\d F_2/\d \ln Q^2$, can provide a more sensitive test of
the $Q^2$ dependence of $F_2$ and hence of the elusive higher twist
effects in deep inelastic scattering.

Finally, we address the issue of the slope of the longitudinal
structure function, $\d F_L/\d\ln Q^2$, in eq.~(2), which has been
neglected in this analysis. We estimate the size of this contribution
using the GRV (1994) parametrisation. We find that this quantity does
not exceed 0.1 for any of the $(x, Q^2)$ values of interest here.
Furthermore, its contribution at small $y$ is suppressed because of
$f_2$; $f_2 < 0.1$ for the small $y$ sample, so that this contribution
never exceeds a percent. For the large $y$ sample, $f_2$ is larger,
$f_2 \sim 0.45$; however, the slope of $F_L$ at these $Q^2$ values is
small enough so that the term contributes less than about 5\%, which is
small compared to the size of the error bars of this data set. Hence it
is reasonable and consistent to ignore the contribution from this
term.

In conclusion, we have analysed a limited sample of the HERA H1 data
\cite{H1sr} on the reduced cross-section, $\sigma_r$, along with
available data on the longitudinal structure function, $F_L$,
\cite{H1fl}, in order to study the $Q^2$ dependence of $\d F_2/\d\ln
Q^2$ {\it in a perturbative regime}. A comparison with available
twist--2 NLO parametrisations shows indications of large $Q^2$
dependences in the data for $\d F_2/\d\ln Q^2$---larger than that
indicated by purely twist--2 behaviour. This is especially true for
small $x$ values ($\sim 10^{-4}$), which also correspond to a smaller
$Q^2 (\sim 4$ GeV${}^2$), indicating that substantial higher twist
effects may be operative here. While the effect is clearly marked only
in the first data point, and the data analysed is obviously limited,
the trend of the data is tantalisingly similar to that expected from
higher twist effects. We therefore urge a detailed analysis of the
$\sigma_r$ data at various $(x, Q^2)$ values in order to shed more
light on the role of higher twists in $F_2$. More data (and improved
errors on $F_L$) will be needed to refine this observation.

\begin{figure}[thp]
\vskip 9truecm
{\includegraphics{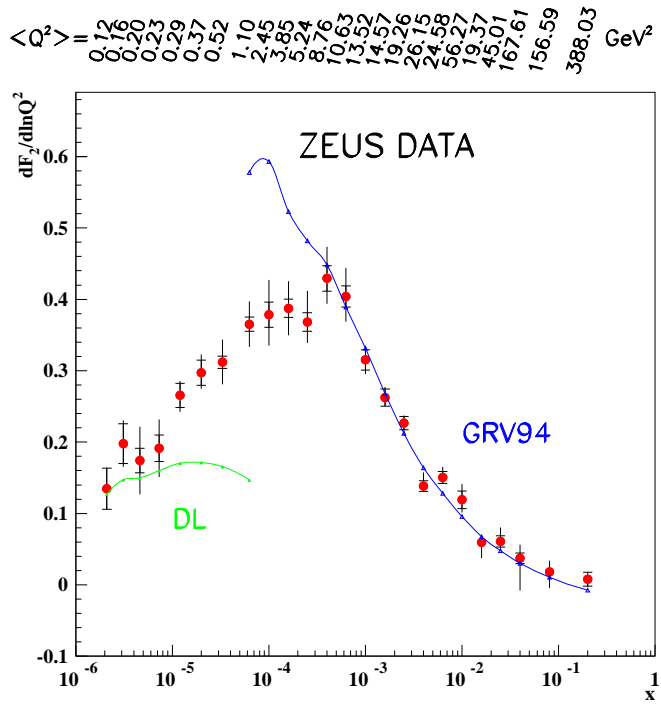}}
\caption[dummy]{Preliminary data on $\d F_2/\d\ln Q^2$ as a function of
$x$ from ZEUS; the graph is taken from reference \cite{Zeusdf2}. The
average $Q^2$ corresponding to each $x$ bin is also shown.}
\end{figure}

\begin{figure}[p]
\vskip 12truecm
{\includegraphics{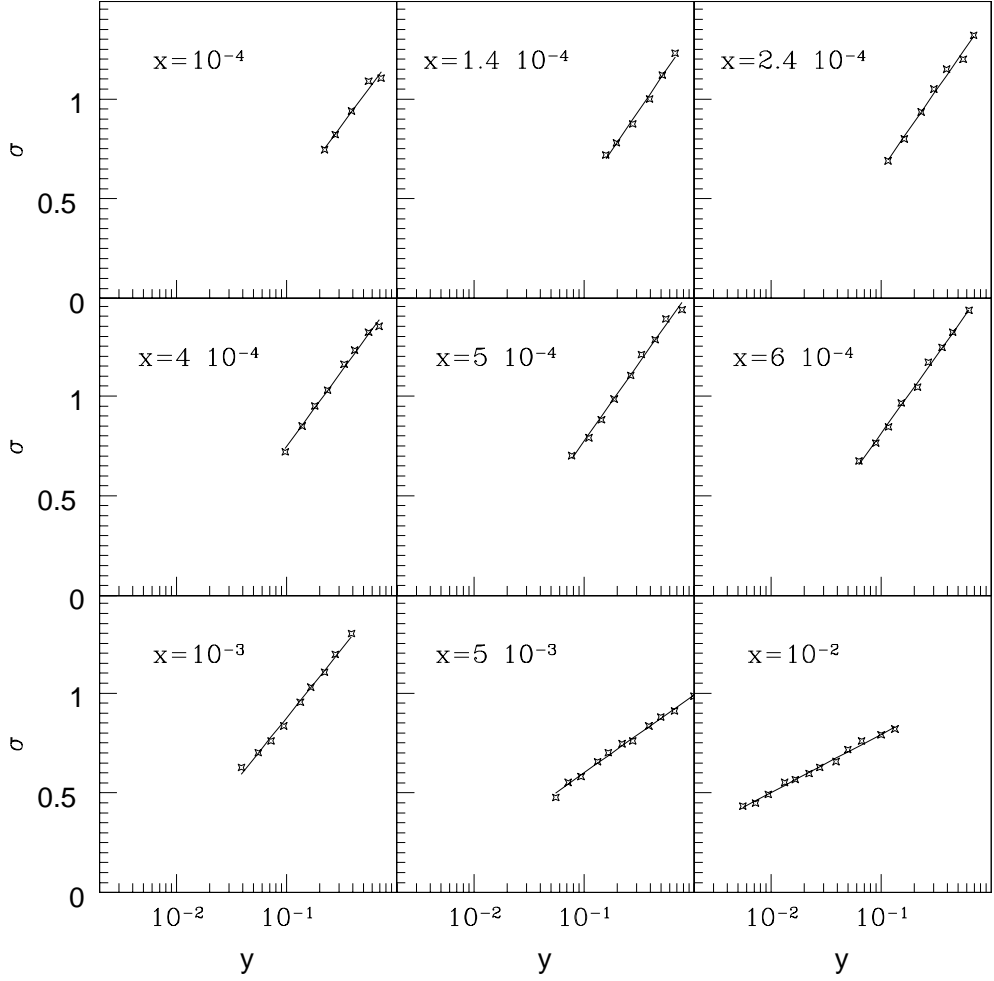}}
\caption[dummy]{The data for the reduced cross-section, $\sigma_r$,
defined in eq.~(1), and taken from \cite{H1sr}, shown as a function of
$y$ along with our straight line fits to $\sigma_r$ as a function of
$\ln y$\  for various $x$ values.}
\end{figure}

\begin{figure}[p]
\vskip 9truecm
{\includegraphics{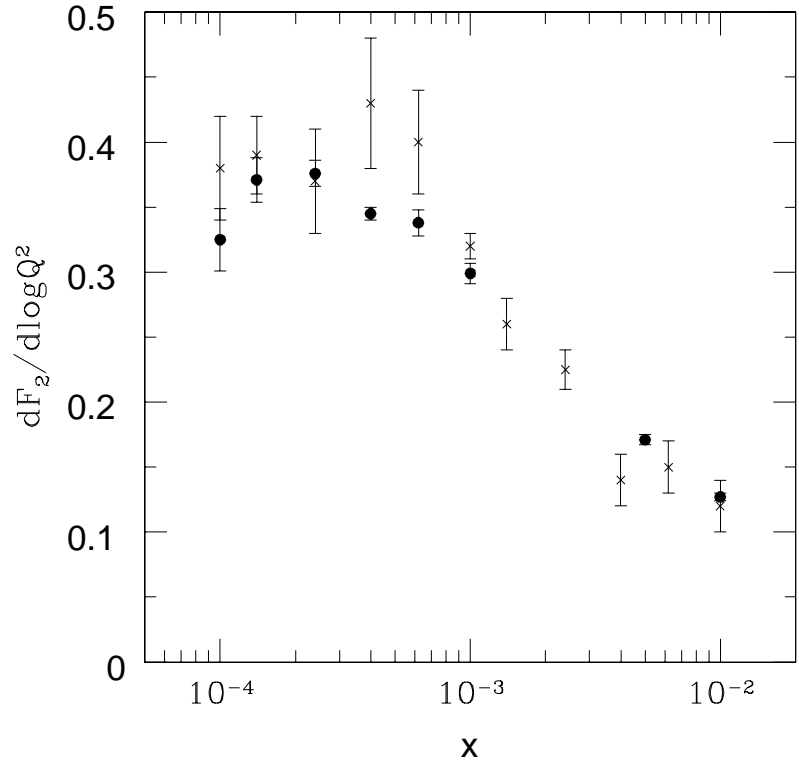}}
\caption[dummy]{The slope $\d F_2/\d\ln Q^2$ as obtained from $\sigma_r$
data (solid circles) as described in the text (small $y$ sample) shown
as a function of $x$ in comparison with ZEUS preliminary data (crosses)
\cite{Zeusdf2}. The $Q^2$ of the points equals the average of that of
the ZEUS data and is shown in Fig.~1.}
\end{figure}

\begin{figure}[p]
\vskip 16truecm
{\includegraphics{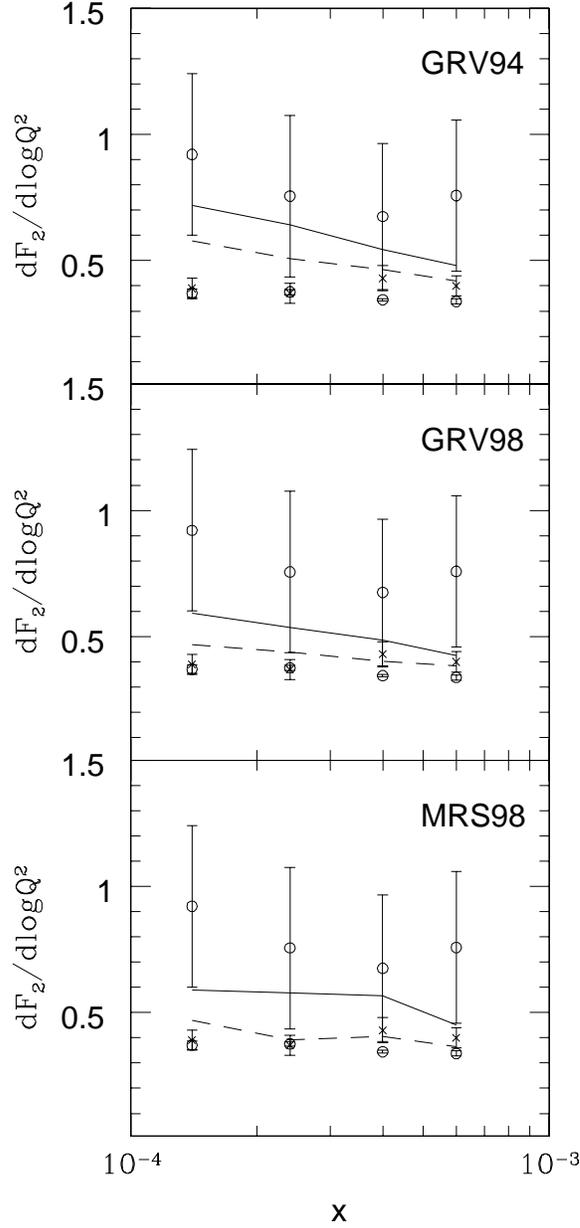}}
\caption[dummy]{The slope $\d F_2/\d\ln Q^2$ (see Table 1) shown as a
function of $x$ in comparison with standard parametrisations,
as obtained from the data given in Refs.~\cite{H1sr,H1fl}
(see text for details).
The upper (lower) circles correspond to the large (small)
$y$ data with the corresponding parametrisations shown as  
solid (dashed) lines. The preliminary ZEUS data \cite{Zeusdf2} (at small
$y$) are also shown (as crosses) in the figure.}
\end{figure}

\end{document}